\documentclass{article}
\usepackage{spconf,amsmath,graphicx}
\usepackage{multirow}
\usepackage{bbding}
\usepackage{amsmath}
\usepackage{amssymb}

\title{SpeechMoE2: MIXTURE-OF-EXPERTS MODEL WITH IMPROVED ROUTING}
\name{Zhao You$^{*1}$, Shulin Feng$^{*1}$, Dan Su$^1$, Dong Yu$^2$}
\address{$^1$Tencent AI Lab, Shenzhen, China\\
$^2$Tencent AI Lab, Bellevue, WA, USA  \\
\{dennisyou, shulinfeng, dansu, dyu\}@tencent.com}
\begin{document}
%
\maketitle
\begin{abstract}
Mixture-of-experts based acoustic models with dynamic routing mechanisms have proved promising results for speech recognition. The design principle of router architecture is important for the large model capacity and high computational efficiency.
Our previous work SpeechMoE only uses local grapheme embedding to help routers to make route decisions. To further improve speech recognition performance against varying domains and accents, we propose a new router architecture which integrates additional global domain and accent embedding into router input to promote adaptability. 
Experimental results show that the proposed SpeechMoE2 can achieve lower character error rate (CER) with comparable parameters than SpeechMoE on both multi-domain and multi-accent task. Primarily, the proposed method provides up to 1.6\% $\sim$ 4.8\% relative CER improvement for the multi-domain task and  1.9\% $\sim$  17.7\% relative CER improvement for the multi-accent task respectively.
Besides, increasing the number of experts also achieves consistent performance improvement and keeps the computational cost constant.  

\end{abstract}

\renewcommand{\thefootnote}{\fnsymbol{footnote}}
\footnotetext[1]{Equal contribution.}
\section{Introduction}

Recently, dynamic mixture-of-experts (MoE) based approaches have been investigated and applied in different tasks such as language modeling \cite{lepikhin2020gshard, fedus2021switch} and image classification \cite{gross2017hard, ahmed2016network, wang2020deep, cai2021dynamic}. Compared with traditional static models, dynamic models can be more efficient in training and inference by selectively activating model components(e.g. layers \cite{huang2017multi}, channels \cite{lin2017runtime}, sub-networks \cite{shazeer2017outrageously}) on conditioned inputs. With this property, they are more adaptable to various inputs, and can also scale up to very large models to better perform while keeping computation cost constant.

our previous work investigated one type of dynamic model in speech recognition, named SpeechMoE \cite{you2021speechmoe}. It has multiple mixture-of-experts layers, using a dynamic routing mechanism to dispatch speech frames among experts networks in each layer. In SpeechMoE, a shared embedding network trained with CTC \cite{graves2006connectionist} criterion is used to provide a high-level representation for routers, helping routers to get a better selective effect. 
However, speech recognition systems need to be robust with different input conditions such as accent, recording channels and acoustic environments in real-world applications. Using the grapheme information from the embedding network for routers, SpeechMoE is not sensitive enough to handle different accents or domains inputs. Explicitly utilizing corresponding information may solve the problem. \cite{li2018multi} has shown that incorporating global dialect-specific information(e.g., one-hot vector of the target dialect) is effective in modeling dialect variations. \cite{sainath2020streaming} adds global domain information to the end-to-end models and shows that feeding the domain-id improves the model performance.

In this study, we continue to explore a mixture-of-experts approach for speech recognition, augmenting SpeechMoE with extra global knowledge to deal with varying domains and accents. The previous SpeechMoE only provides frame-level grapheme embedding for routers. Here, we propose a new router architecture which takes extra domain embedding and accent embedding as router input. To constrain the computation cost of the embedding network, we use multi-task training for the original embedding network to simultaneously produce grapheme, domain, and accent embedding. 
We hypothesize that explicitly providing domain and accent embedding for routers should help improve the route decisions, making the experts specialized in processing distinct input utterances. 
Besides, we use FastMoE \cite{he2021fastmoe} in our system to support scaling up the MoE model size by training across multiple GPUs on multiple nodes. With comparable computation costs, more experts also help achieve better performance.

The rest of the paper is organized as follows. Section 2 reviews the previous work of SpeechMoE, and Section 3 presents our proposed method SpeechMoE2. The experimental setup is described in Section 4, and the experimental results are reported in Section 5. Finally, we conclude this paper in Section 6. 

\label{sec:typestyle}
\begin{figure*}[!tb]
\begin{minipage}[b]{1.0\linewidth}
  \centering
  \centerline{\includegraphics[width=14.5cm]{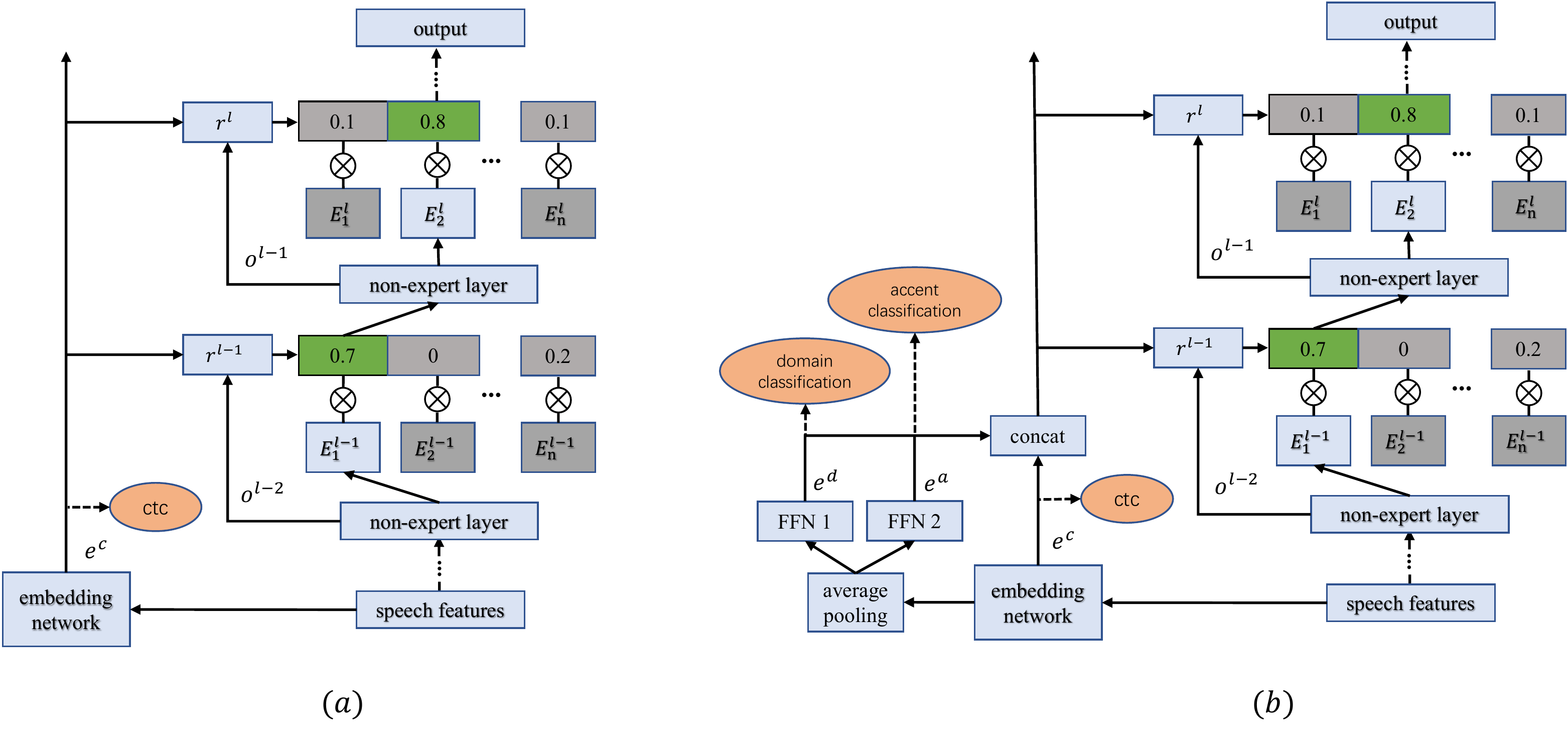}}
\end{minipage}
\vspace{-15pt}
\caption{
(a) and (b) illustrate the architecture of SpeechMoE and SpeechMoE2 respectively. In SpeechMoE2, multi-task training is conducted on the embedding network to learning grapheme embedding, domain embedding, and accent embedding simultaneously, which will be concatenated as extra knowledge to feed routers.
}
\vspace{-10pt}
\end{figure*}

\section{Previous works}
\label{sec:format}

In this section, we mainly describe our previous work of SpeechMoE.

\subsection{SpeechMoE}
The SpeechMoE is proposed in \cite{you2021speechmoe}, and it can achieve higher prediction accuracy than traditional static models with comparable computational costs of inference.

The architecture of SpeechMoE is described in Fig.1(a), which comprises multiple MoE layers, non-expert layers and a shared embedding network. Each MoE layer consists of $n$ experts and a router layer. It takes the output of the previous layer
and the shared embedding as input and routes each speech frame to the top-1 expert with the largest route probability. Let $W^{l}_{r}$, $e^{c}$ and $o^{l-1}$ be the router weights of the $l$-th layer, shared embedding and the output of the previous layer, then the router probability can be defined as follows:




\begin{equation}
  r^{l} = W_{r}^{l} \cdot Concat(e^{c};o^{l-1})
\end{equation}

\begin{equation}
  {p^{l}_{i}} = \frac{exp^{r^{l}_{i}}}{\sum_{j=1}^{n}exp^{r^{l}_{j}}}
\end{equation}
Then, the selected expert's output is also gated by router probability to get the output of the MoE layer,
\begin{equation}
y^{l} = p^{l}_{i} E^{l}_{i}
\end{equation}

Since only one expert is active in each layer, the SpeechMoE can keep the computational cost constant while scaling up to a very large model. To achieve better sparsity and balance among different experts, the sparsity L1 loss $L_{s}$ and mean importance loss $L_{m}$ are added into the loss function:

\begin{equation}
L_{s}= \frac{1}{k}\sum_{j=1}^{k} \parallel{\hat{p}_{j}}\parallel_{1}
\end{equation}

\begin{equation}
{L}_{m}= n\sum_{i=1}^{n} ({\frac{1}{k}\sum_{j=1}^{k} p_{ij}})^{2}
\end{equation}
where $\hat{p}_{j}$ stands for the unit-normalized router probability distribution of frame $j$, and $k$ is the number of frames in this mini-batch. $p_{ij}$ stands for the router probability on expert $i$ of frame $j$.

The loss function for training SpeechMoE is defined as:
\begin{small}
\begin{equation}
\begin{split}
L(x;y)=L_{c}(x;y) +\alpha L_{s}(x) + \beta {L}_{m}(x) + \gamma  L_{e}(x;y)
\end{split}
\end{equation} 
\end{small}
where $x$ and $y$ are the input speech features and target label, respectively. $L_{c}$ is the CTC loss for speech recognition, while $L_{s}$ and ${L}_{m}$ are the mentioned sparsity $L1$ loss and mean importance loss, used to encourage sparsity and diversity of the model. The embedding loss $L_{e}$ is also CTC loss. It shares the same goal with our SpeechMoE model and provides reliable embedding for the routers. $\alpha$, $\beta$, and $\gamma$ are the scale for $L_{s}$, ${L}_{m}$ and $L_{e}$ respectively.

\section{SpeechMoE2}
\subsection{Model structure}
Fig.1(b) illustrates the architecture of the proposed SpeechMoE2. Its main structure is the same as the SpeechMoE while the difference is that the embedding network is revised to output domain and accent embedding besides grapheme embedding. In order not to significantly increase the computation cost, we simply add an average pooling layer and two feed-forward layers over the embedding network to do multi-task training on domain and accent classification. Then, embedding from three aspects of the domain, accent and CTC will be concatenated with the previous layer's output as the input of routers, which can be defined as:


\begin{equation}
r^{l} = W_{r}^{l} \cdot Concat(e^{c};e^{a};e^{d};o^{l-1})
\end{equation}

\begin{equation}
e^{a} = W_{a} \cdot \frac{1}{T}\sum_{t=1}^{T} e_{t}^{c}
\end{equation}

\begin{equation}
e^{d} = W_{d} \cdot \frac{1}{T}\sum_{t=1}^{T} e_{t}^{c}
\end{equation}
Where $W_{a}$ and $W_{d}$ represent projection weight for accent and domain respectively. $T$ is the total frames number for a certain utterance.

\begin{table}
\caption{\textit{Configurations for different types of test sets} }
\vspace{5pt}
\label{tab:1}
\begin{center}
\scalebox{0.85}{
    \begin{tabular}{|c|c|c|c|c|}
    \hline
    \multirow{1}*{{Data type}} &
    \multirow{1}*{{vp}} & 
    \multirow{1}*{\small{rir}} & 
    \multirow{1}*{\small{aec}} &
   \multirow{1}*{\small{noise}}  \\
  \hline
    \small{CL}    & ×  & ×      & ×  & ×  \\
    \small{NVR}      & × &  \checkmark & ×  & × \\
    \small{VR}      & \checkmark & \checkmark    & ×  & × \\
    \small{VRA}      & \checkmark & \checkmark    & \checkmark  & × \\
    \small{VRN}      & \checkmark & \checkmark    & ×  & \checkmark \\
    \small{VRNA}      & \checkmark & \checkmark    & \checkmark  & \checkmark \\
     \hline
    \end{tabular}
}
\end{center}
\vspace{-10pt}
\end{table}

Given the input $x$, grapheme target $y$, domain target $y_{d}$ and accent target $y_{a}$, the full loss function of our method is defined as 
\begin{small}
\begin{equation}
\begin{split}
L(x;y)=L_{c}(x;y) +\alpha L_{s}(x) + \beta {L}_{m}(x) \\ 
+ \gamma  L_{e}(x;y) + \eta  L_{a}(x;y_{a}) + \theta L_{d}(x;y_{d})
\end{split}
\end{equation} 
\end{small}Among these items, $L_{c}$, $L_{s}$, ${L}_{m}$ and $L_{e}$ are the same as mentioned above. $L_{a}$ and $L_{d}$ are the accent and domain classification loss.  $\alpha$, $\beta$, $\gamma$, $\eta$, and $\theta$ are the scale for $L_{s}$, ${L}_{m}$, $L_{e}$, $L_{a}$ and $L_{d}$ respectively.

\subsection{Toward Large models}
We use FastMoE \cite{he2021fastmoe} to support scaling up the model size by training across multiple GPUs on multiple nodes, where experts are placed on different devices while non-expert layers and routers are copied. Different workers process different batches of data and speech frames in them will communicate across experts by the route decisions. We have introduced the mentioned $L_{s}$ and $L_{m}$ loss to make the router probabilities sparse and diverse. Since large batch size will lead to large route decision space, we collect all router probabilities from different workers and compute the complete loss and gradients, which would be equivalent to dealing with a larger data batch and may help improve model performance.

\begin{table}
\caption{\textit{Total length (approx.hours) of each domain for training and testing (k for thousand). } }
\label{tab:1}
\begin{center}
\scalebox{0.85}{
    \begin{tabular}{|c|c|c|c|c|c|c|}
    \hline
    \multirow{1}*{{Model}} &
    \multirow{1}*{{CL}} &
    \multirow{1}*{{NVR}} &
    \multirow{1}*{{VR}} &
    \multirow{1}*{{VRA}} &
    \multirow{1}*{{VRN}} &
    \multirow{1}*{{VRNA}} \\
  \hline
    \small{Train}    & 89k & 7k & 30k & 8k & 8k & 8k \\
    \hline
     \small{Test}    & 2.1 & 2.1 & 2.1 & 2.1 & 2.1 & 2.1 \\

     \hline
    \end{tabular}
}
\end{center}
\vspace{-10pt}
\end{table}

\begin{table}
\caption{\textit{Total length (approx.hours) of each accent (k for thousand). } }
\vspace{-10pt}
\label{tab:1}
\begin{center}
\scalebox{0.85}{
    \begin{tabular}{|c|c|c|c|c|c|c|}
    \hline
    \multirow{1}*{{Accent}} &
    \multirow{1}*{{sichuan}} &
    \multirow{1}*{{hunan}} &
    \multirow{1}*{{fujian}} &
    \multirow{1}*{{guangdong}} \\
  \hline
    \small{Train}    & 3.2k & 4.5k & 4.3k & 4.5k \\
    \hline
     \small{Test}    & 4.1 & 4.3 & 4.4 & 4.0 \\

     \hline
    \end{tabular}
}
\end{center}
\vspace{-10pt}
\end{table}

\section{Experimental Setup}
\label{sec:format}

\subsection{Datasets}

Our training corpus is mixed data sets collected from several different application domains and accents.  
It comes to a 150k hours training corpus. In this work, we use the term ‘domain’ to mean a logical group of utterances that share similar characteristics such as background noise. 


To evaluate the performance of augmenting with domain embedding, we have simulated six types of domain test sets. They are constructed with different mixing conditions like voice processing(vp), acoustic echo cancellation(aec), reverberation(rir) and signal-to-noise-ratio(SNR), etc. The SNR is set between 15 and 30 dB, while the reverberation time is set between 0 and 900 milliseconds. We refer these test sets as clean(CL), nonvp-rir(NVR), vp-rir(VR), vp-rir-noise(VRN), vp-rir-aec(VRA), vp-rir-aec-noise(VRAN) respectively. The detailed configurations for them are shown in Table 1 and the amount of domain-specific training and test data can be found in Table 2.


The training utterances include four different accents, namely sichuan, hunan, fujian and guangdong. The amount of training and test data is shown in Table 3.



\subsection{Training setup}

 All the experiments use 40-dimension log-mel filterbank features appended with the first-order and the second-order derivatives. Log-mel filterbank features are computed with a 25ms window and shifted every 10ms. These features are stacked with 8 consecutive frames and downsampled to a 30ms frame rate. Each frame uses a global mean and variance normalization. All networks are trained with the CTC criterion.  We use the context-independent-syllable-based acoustic modeling method \cite{syllable} for CTC learning. The target labels of CTC learning are defined to include 1394 Mandarin syllables, 39 English phones, and a blank. Character error rate results are measured on the test sets, and the floating-point operations (FLOPs) for a one-second example are used to evaluate the computational cost of inference. We use a pruned, first pass, 5-gram language model for decoding. All the systems use a vocabulary that consists of millions of words. Decoding is performed with a beam search algorithm by using the weighted finite-state transducers (WFSTs).


\subsection{Acoustic Model}

The backbone of our model consists of 30 MoE layers, 30 sequential memory layers and 3 self-attention layers. Each MoE layer is followed by one sequential memory layer, and a self-attention layer is inserted after each 10 consecutive MoE and sequential memory layers. The detailed configuration of each layer is consistent with \cite{you2021speechmoe}. In our experiments, we vary the number of experts of MoE layers to be 2, 4 and 16, which are marked as MoE-2e, MoE-4e and MoE-16e respectively. The shared embedding network is a static model without MoE layers but a similar structure to the backbone. Additionally, accent and domain projection layers are added to the shared embedding network.  For the accent and domain projection layers, we set the embedding dimension to be 512.


In our study, we built the baseline SpeechMoE system (MoE1) for evaluating the performance of our proposed method.
The baseline model has 2 experts in each MoE layer, not using the domain and accent embedding.
For all experiments on MoE models, we set the hyper-parameters $\alpha=0.05$, $\beta=0.05$, $\gamma=0.01$, $\eta=0.1$ and $\theta=0.1$.

\newcommand{\tabincell}[2]{\begin{tabular}{@{}#1@{}}#2\end{tabular}} 
\begin{table}
\caption{\textit{Results of augmenting with domain embedding. } }
\label{tab:1}
\begin{center}
\scalebox{0.71}{
\begin{tabular}{|c|c|c|c|c|c|c|c|c|}
\hline
\multirow{2}*{\small{Model}} & \multirow{2}*{\small{Params}} & \multirow{2}*{\small{FLOPs}} & \multicolumn{6}{c|}{\small{Test set}} \\
\cline{4-9}
&  &  & \small{CL} & \small{NVR} & \small{VR} &  \small{VRA} &  \small{VRN} & \small{VRNA}  \\
\hline
\small{MoE1}    & 105M     & 2.3B & 14.09 & 17.64 & 18.22  & 26.68 &  22.43  & 33.67 \\
\small{MoE2-2e}    & 107M    & 2.3B & \textbf{13.55} & \textbf{17.34} & \textbf{17.87}  & \textbf{26.07} & \textbf{21.36} & \textbf{33.13} \\
 \hline
\end{tabular}
}
\end{center}
\vspace{-5pt}
\end{table}



\begin{table}
\caption{\textit{Results of augmenting with accent embedding. } }
\label{tab:1}
\begin{center}
\scalebox{0.83}{
\begin{tabular}{|c|c|c|c|c|c|c|c|c|}
\hline
\multirow{2}*{\small{Model}} & \multirow{2}*{\small{Params}} & \multirow{2}*{\small{FLOPs}} & \multicolumn{4}{c|}{\small{Test set}} \\
\cline{4-7}
&  &  & \small{sichuan} & \small{guangdong} & \small{fujian} &  \small{hunan} \\
\hline
\small{MoE1}    & 105M     & 2.3B & 12.81 & 7.99 & 6.59  & 34.25 \\
\small{MoE2-2e}    & 107M    & 2.3B & \textbf{10.54} & \textbf{7.84} & \textbf{6.39}  & \textbf{31.00}  \\
 \hline
\end{tabular}
}
\end{center}
\vspace{-15pt}
\end{table}




\section{Experimental Results}
\subsection{Augmenting with extra knowledge}
In this section, we evaluate the performance of augmenting with domain and accent embedding. As shown in Table 4, the MoE2-2e, which uses extra domain embedding and accent embedding, achieves lower character error rate than the baseline model for the multi-domain task. Similar results can be seen in Table 5, MoE2-2e can also achieve lower character error than the baseline model for the multi-accent task.

\begin{table}
\caption{\textit{Results of increasing the number of experts for multi-domain task. } }
\vspace{-5pt}
\label{tab:1}
\begin{center}
\scalebox{0.71}{
\begin{tabular}{|c|c|c|c|c|c|c|c|c|}
\hline
\multirow{2}*{\small{Model}} & \multirow{2}*{\small{Params}} & \multirow{2}*{\small{FLOPs}} & \multicolumn{6}{c|}{\small{Test set}} \\
\cline{4-9}
&  &  & \small{CL} & \small{NVR} & \small{VR} &  \small{VRA} &  \small{VRN} & \small{VRNA}  \\
\hline
\small{MoE2-2e}    & 107M    & 2.3B & 13.55 & 17.34 & 17.87  & 26.07 & 21.36 & 33.13 \\
\small{MoE2-4e}    & 171M    & 2.3B & 13.21 & 17.07 & 17.41  & 25.41 & 21.05 & 32.54 \\
\small{MoE2-16e}    & 500M    & 2.3B & \textbf{12.79} & \textbf{16.82} & \textbf{16.96}  & \textbf{24.45} & \textbf{20.12} & \textbf{31.46} \\
 \hline
\end{tabular}
}
\end{center}
\vspace{-5pt}
\end{table}

\begin{table}
\caption{\textit{Results of increasing the number of experts for multi-accent task. } }
\vspace{-5pt}
\label{tab:1}
\begin{center}
\scalebox{0.83}{
\begin{tabular}{|c|c|c|c|c|c|c|c|c|}
\hline
\multirow{2}*{\small{Model}} & \multirow{2}*{\small{Params}} & \multirow{2}*{\small{FLOPs}} & \multicolumn{4}{c|}{\small{Test set}} \\
\cline{4-7}
&  &  & \small{sichuan} & \small{guangdong} & \small{fujian} &  \small{hunan} \\
\hline
\small{MoE2-2e}    & 107M     & 2.3B & 10.54 & 7.84 & 6.39  & 31.00 \\
\small{MoE2-4e}    & 171M    & 2.3B & 9.16 & 7.73 & 6.31  & 29.63  \\
\small{MoE2-16e}    & 500M    & 2.3B & \textbf{8.92} & \textbf{7.59} & \textbf{6.24}  & \textbf{28.49}  \\
 \hline
\end{tabular}
}
\end{center}
\vspace{-5pt}
\end{table}

The experimental results indicate that providing a high-level representation for routers to make them aware of the varying domains and accents would be helpful to get a better route strategy and improve the performance. But, more importantly, the proposed architecture may prove to be a flexible framework to integrate diverse global information to adapt different tasks.

\subsection{Increasing the number of experts}
In this section, we investigate the effect of increasing the number of experts. 
Table 6 and Table 7 show the performance of SpeechMoE2 models with different number of experts on multi-domain and multi-accent tasks. 
The results show that the performance of SpeechMoE2 gets better consistently as the number of experts increases. From 2 experts to 16 experts, the relative gain is averaged 5.1\% for domain-specific test sets and 7.7\% for accent-specific test sets. Specially, as we increase the number of experts,  our models keep the FLOPs per frame constant.






\section{Conclusions and future work}
In this paper, we extend our previous work to improve the performance of multi-domain and multi-accent speech recognition. The improvement comes from the new router architecture, which takes extra domain embedding and accent embedding as input. The idea is intuitive that providing extra knowledge can make the router more specialized for varying input utterances and experimental results prove the effectiveness. For multi-domain task, the proposed method provides up to 1.6\% $\sim$ 4.8\% relative CER improvement. For the multi-accent task, the relative gain is between 1.9\% $\sim$  17.7\%. Besides, increasing the number of experts also achieves consistent performance improvement. In the future, we plan to explore new MoE layer with more complicated model such as conformer.




\bibliographystyle{IEEEbib}
\bibliography{strings,refs}

\end{document}